\documentclass{JHEP3}
\usepackage{graphicx}
\def\be{\begin{eqnarray}}\def\ba{\begin{eqnarray}}
\def\ee{\end{eqnarray}}\def\ea{\end{eqnarray}}

\title{Hybrid exotic meson with $J^{PC}=1^{-+}$ in AdS/QCD}

\author{Hyun-Chul Kim \\ Department of Physics,
Inha University, Incheon 402-751, Korea \\
E-mail: \email{hchkim@inha.ac.kr} }
\author{Youngman Kim \\ School of Physics, Korea Institute for
  Advanced
Study, Seoul 130-012, Korea \\
E-mail: \email{chunboo81@kias.re.kr}}

\abstract{We investigate the hybrid exotic meson with $J^{PC}=1^{-+}$
within the framework of an AdS/QCD model.  Introducing a holographic
field dual to the operator for hybrid exotic meson, we obtain the
eigen-value equation for its mass. Fixing all free parameters by
QCD observables such as the $\rho$-meson mass, we predict the masses
of the hybrid exotic meson.  The results turn out to be
$1476\,\mathrm{MeV}$ for the ground state, and $2611\,\mathrm{MeV}$
for the first excited one. Being compared with the existing
experimental data for the $\pi_1(1400)$, which is known to be
$m_{\pi_1}\;=\;1351\pm30\,\mathrm{MeV}$, the present result seems to
be qualitative in agreement with it.  We also predict the decay
constant of $\pi_1$(1400): $F_{\pi_1}= 10.6$ MeV.
}

\preprint{INHA-NTG-14/2008}

\begin{document}

\section{Introduction}
There has been a considerable amount of interest in exotic mesons
well over decades, since it cannot be explained by the conventional
constituent quark model in which mesons consist of quark and antiquark
pairs ($q\bar{q}$).  The nonexotic mesons are restricted to have
quantum numbers constrained by the following selection rules:
$P=(-1)^{L+1}$ and $C=(-1)^{L+S}$, where $L$ and $S$ denote the
relative orbital angular momentum and total spin of quarks
consisting of mesons. Thus, an exotic meson with
$J^{PC}=1^{-+}$ cannot be explained as a $q\bar{q}$ state.
It is only possible to produce such a state either as a multi-quark
state (tetraquark) or as a quark-gluon hybrid state.  In particular,
it is of great importance to understand the quark-gluon hybrid exotic
mesons, since it provides a key to examine the role of the
gluons as basic building blocks for hadrons.

Since Jaffe and Johnson suggested a possible existence of hybrid
mesons~\cite{Jaffe:1975fd}, there has been a great deal of theoretical
investigations~(see a recent review~\cite{Klempt:2007cp} for full
references): For example, the bag
model~\cite{Barnes:1977hg,deViron:1980zf,Barnes:1982zs,
Chanowitz:1982qj,Barnes:1982tx}, the flux tube
model~\cite{Isgur:1985vy,Close:1994hc,Barnes:1995hc,Page:1998gz,Burns:2006wz},
the QCD sum
rules~\cite{Balitsky:1982ps,Balitsky:1986hf,Latorre:1985tg,
Govaerts:1984bk,Govaerts:1984hc,Govaerts:1986pp},
lattice QCD~\cite{Bernard:2003jd,Hedditch:2005zf,McNeile:2006bz},
etc.  Experimentally, a hint of the exotic meson was already observed
many years ago~\cite{Apel:1981wb}.  Later, the lowest-lying hybrid
exotic meson has been found by various experimental
collaborations~\cite{Thompson:1997bs,Chung:1999we,Adams:2006sa,Abele:1998gn,
Abele:1999tf,Amelin:2005ry} and was christened $\pi_1(1400)$.
The $\pi_1(1400)$ is now announced with its mass
$m_{\pi_1}\;=\; 1351\pm 30$ MeV and width $\Gamma_{\pi_1}\;=\;313\pm
40$ MeV by the Particle Data Group (PDG)~\cite{Amsler:2008zz}.

The AdS/CFT
correspondence~\cite{Maldacena:1997re,Gubser:1998bc,Witten:1998qj}
that connects a strongly coupled large $N_c$ gauge theory to a weakly
coupled supergravity provides novel and attractive insight into
nonperturbative features of quantum chromodynamics (QCD) such as the
quark confinement and spontaneous breakdown of chiral symmetry
(SB$\chi$S).  Though there is still no rigorous theoretical ground for
such a correspondence in real QCD, this new idea has triggered a great
amount of theoretical works on possible mappings from nonperturbative
QCD to $5D$ gravity, i.e. holographic dual of QCD.  In fact, there are
in general two different routes to modeling holographic dual of QCD
(See, for example, a recent review~\cite{Erdmenger:2007cm}): One way
is to
construct 10 dimensional (10D) models based on string theory of D3/D7,
D4/D6 or D4/D8 branes~\cite{Karch:2002sh,Kruczenski:2003be,
Kruczenski:2003uq,Sakai:2004cn,Sakai:2005yt}.  The other way is
so-called a bottom-up approach to a holographic model of
QCD, often called as AdS (Anti-de Sitter
Space)/QCD~\cite{Erlich:2005qh,DaRold:2005zs,DaRold:2005vr} in which a
5D holographic dual is constructed from QCD, the 5D gauge coupling
being identified by matching the two-point vector correlation
functions.  Despite the fact that this bottom-up approach is somewhat
on an ad hoc basis, it reflects some of most important features of
gauge/gravity dual. Moreover, it is rather successful in describing
properties of hadrons (See the recent
review~\cite{Erdmenger:2007cm}).

In the present work, we want to investigate the quark-gluon hybrid
meson with $J^{PC}=1^{-+}$, in particular, $\pi_1(1400)$, based on the
AdS/QCD model developed by
refs.~\cite{Erlich:2005qh,DaRold:2005zs,DaRold:2005vr}.  Since the
hybrid exotic mesons exist in large $N_c$ QCD as narrow resonant
states~\cite{Cohen:1998jb}, it is worth while to study it within
AdS/QCD.  Moreover, the AdS/QCD model has a virtue in dealing with
gluonic operators for ease of application, since the 5D bulk fields
corresponding to them can be much more easily handled.
In AdS/QCD, the $\pi_1(1400)$ may be regarded as a spin-1 bulk field
with quantum number $J^{PC}=1^{-+}$. It can be identified as a first
excitation of the Kaluza-Klein (KK) modes in this channel.  The
eigenvalue equation for the $\pi_1(1400)$ is derived from the
classical equation of motion from the 5D action, so that the mass
of the $\pi_1(1400)$ can be obtained by solving this equation.  In
fact, we will show that the $\pi_1(1400)$ mass arises from the first
zero of the modified Bessel function $K_3$.

The present work is organized as follows: In section 2, we briefly
review the present status of theoretical works for the hybrid exotic
mesons. In section 3, we explain briefly the hard-wall AdS/QCD model
with the bulk field for the hybrid exotic meson taken into account.
Then, we present the result for the mass of the hybrid exotic meson
with $J^{PC}=1^{-+}$ and compare it with those of various models and
experimental data.  We also show the result of the decay constant of
the $\pi_1$ hybrid exotic meson.  In the last section, we summarize
the present work.

\section{Hybrid exotic meson with $J^{PC}=1^{-+}$}
As already mentioned in Introduction, the hybrid exotic mesons have
been studied extensively in many different theoretical frameworks.  In
1980s, various versions of the MIT bag model with transverse gluon
fields were invoked to predict the existence of the hybrid exotic meson
with $J^{PC}=1^{-+}$ under the name of $q\bar{q}G$ hermaphrodite meson
or meikton~\cite{Barnes:1977hg,deViron:1980zf,Barnes:1982zs,
Chanowitz:1982qj,Barnes:1982tx}.  In the MIT bag model, the lowest
state of hybrid exotic mesons consists of $(q\bar{q})$ and of a
transverse TE (magnetic) gluon that is the lowest gluonic eigenmode
due to the MIT boundary conditions.  Having considered
$\mathcal{O}(\alpha_{\mathrm{s}})$-order energy shifts,
refs.~\cite{Chanowitz:1982qj,Barnes:1982tx} predicted the mass of the
lowest hybrid exotic state with $J^{PC}=1^{-+}$ to be around $1400$
MeV.

The flux tube model was frequently used to investigate the hybrid
exotic mesons.  The model was extracted from the strong coupling
Hamiltonian lattice formulation of QCD~\cite{Isgur:1984bm}.  The term,
``flux tube'', mimics a roughly cylindrical region of chaotic gluon
fields, which confines widely separated static color sources.  This
flux tube leads to a confining linear potential between color singlet
$q$ and $\bar{q}$.  The model contains normal modes of excitation only
to the locally transverse spatial direction.  The model predicts the
lowest state of hybrid exotic meson in the range of $1800-2100$
MeV~\cite{Isgur:1985vy,Close:1994hc,Barnes:1995hc,Page:1998gz,
Swanson:2003kg,Burns:2006wz}.

In QCD sum rules, the predictions of the mass of the hybrid exotic
meson with $J^{PC}=1^{-+}$ do not seem consistent: For example,
Balitsky et al.~\cite{Balitsky:1982ps,Balitsky:1986hf} estimate the
mass around $1000-1300$ MeV, while Latorre et
al.~\cite{Latorre:1985tg} vote for $M(1^{-+})\approx 2.1$ GeV.
Refs.~\cite{Govaerts:1984bk,Govaerts:1984hc,Govaerts:1986pp} suggest
even $M(1^{-+})\approx 2500$ MeV.

While lattice QCD is the most promising way of describing low-energy
phenomena in QCD, it is still far from the real world, since the pion
mass in lattice QCD is still larger than the physical pion mass
$m_{\pi}=139.57$ MeV.  In fact, Thomas and
Szczepaniak~\cite{Thomas:2001gu} examined chiral extrapolations in
exotic meson spectrum and found that the linear extrapolation does not
seem working, since the self-energy corrections to the exotic meson
mass are most likely to introduce some non-linearity in the chiral
extrapolation of lattice calculation of its mass.

In QCD, the hybrid exotic meson with $J^{PC}=1^{-+}$ can be treated
as a vector operator consisting of the quark, antiquark and gluon:
\begin{equation}
J_\mu^a(x) \;=\; \bar{\psi}(x) T^a G_{\mu\alpha}(x)\gamma^\alpha \psi(x),
\label{eq:vector}
\end{equation}
where $\psi(x)$ and $G_{\mu\alpha}$ denote the quark field and the gluon
field strength defined as $G_{\mu\alpha} \;=\; G_{\mu\alpha}^A t^A$ with
color matrices $t^A$ ($\mathrm{tr}[t^A t^B]=\delta_{AB}/2$). The $T^a$
represent the flavor matrices and we take it as Pauli matrices, since
we consider only flavor SU(2) in the present work.
The two-point correlation function~\cite{Balitsky:1986hf} for the
vector current in eq.~(\ref{eq:vector}) is written as
\begin{equation}
  \label{eq:corr1}
\Pi_{\mu\nu} \;=\; \int d^4 x e^{iq\cdot x} \langle
T(J_\mu^a(x)J_\nu^b(0))\rangle_0 \;=\;
-\left(g_{\mu\nu} -\frac{q_\mu q_\nu}{q^2}\right) \Pi_V^{ab}(q^2) +
\frac{q_\mu q_\nu}{q^2} \Pi_S^{ab}(q^2),
\end{equation}
where $\Pi_V(q^2)$ includes the intermediate hybrid
exotic vector mesons with $J^{PC}=1^{-+}$, whereas $\Pi_S(q^2)$
contains the hybrid exotic scalar mesons $J^{PC}=0^{++}$. The result
of the operator product expansion (OPE)~\cite{Balitsky:1986hf} is
given as follows:
\begin{eqnarray}
\Pi_V (q^2) &=& -\frac{(N_c^2-1)}{2^7 15\pi^3} q^6 \ln(-q^2) +
\cdots,\label{vOPE}\\
\Pi_S (q^2) &=& -\frac{(N_c^2-1)}{2^8 15\pi^3} q^6 \ln(-q^2) + \cdots,
\label{eq:corr2}
\end{eqnarray}
where $\Pi_{V,S}\delta^{ab}/2 = \Pi_{V,S}^{ab}$.

We will use this expression (\ref{eq:corr2}) to fix the 5D bulk field
for the hybrid exotic meson.

\section{Hybrid exotic mesons in AdS/QCD}
The metric of an AdS space is defined as
\begin{equation}
ds^2\;=\;g_{MN}dx^M dx^N \;=\;\frac1{z^2}(\eta_{\mu\nu} dx^\mu dx^\nu -
dz^2),
\end{equation}
where $\eta_{\mu\nu}$ denotes the 4D Minkowski metric:
$\eta_{\mu\nu}=\mathrm{diag}(1,\,-1,\,-1,\,-1)$. The AdS space is
compactified:  IR
boundary at $z=z_m$, and  UV at $z=\epsilon\to 0$. Thus, the model
is justified within the range: $\epsilon\le z \le z_m$.   The 5D
gauge action in $\mathrm{AdS}_5$ space with the bulk field for the
hybrid exotic meson with $J^{PC}=1^{-+}$ contained can be expressed as
\begin{equation}
S\;=\;
\int\,d^4x dz\,\sqrt{g}\,\mathrm{Tr}\left[-\frac1{4\tilde g_5^2}
  F_{MN}F^{MN} +\frac12 m_5^2 V_M V^M \right] ,\label{Act}
\end{equation}
where $F_{MN}\;=\; \partial_M V_N -\partial_N V_M -i[V_M,\,V_N]$ and
$M,\, N=0,1,2,3,4$. The gauge field, which is dual to the 4D current
in Eq. (\ref{eq:vector}), is defined as $V_M =V_M^{a}t^a$ with
$\mathrm{tr}(t^at^b)=\delta^{ab}/2$.  Note that the 5D gauge coupling
$\tilde g_5$ is not identical to the gauge coupling
in~\cite{Erlich:2005qh,DaRold:2005zs}, where the bulk gauge field is
dual to a 4D vector current $\bar\psi (x)\gamma^\mu t^a\psi(x)$.  The
5D mass $m_5^2$ of the bulk field $V_M$ is determined by the relation
$m_5^2\;=\;(\Delta-p)(\Delta+p-4)$~\cite{Gubser:1998bc,Witten:1998qj},
where $\Delta$ stands for the dimension of the corresponding operator
with spin $p$.  Since the dimension and spin of the operator in
Eq.~(\ref{eq:vector}) are $\Delta=5$ and $p=1$, respectively,
we get the 5D mass of the bulk field $V_M$ is: $m_5^2\;=\;
(5-1)(5+1-4)= 8$.  The hybrid exotic meson $\pi_1(1400)$ may be
regarded as a first Kaluza-Kleein (KK) excitation.

In ref.~\cite{Klebanov:1999tb}, it was shown that for small $z$ or
near the boundary of AdS space, a 5D field $\phi(x,z)$ dual to a 4D
operator $\mathcal{O}$ can be expressed as
\begin{equation}
\phi(x,z)\;=\; z^{4-\Delta-p}  \left[\phi_0(x) + O(z^2)\right]+z^{\Delta-p}
\left[A(x) + O(z^2)\right],
\end{equation}
where $\phi_0(x)$ is a prescribed source function for $\mathcal{O}(x)$
and $A(x)$ denotes a physical fluctuation that can be
determined from the source by solving the classical equation of
motion.  It can be directly related to the vacuum expectation value
(VEV) of the $\mathcal{O}(x)$ as follows~\cite{Klebanov:1999tb}:
\begin{equation}
A(x)\;=\;\frac1{2\Delta -4} \langle\mathcal{O}(x)\rangle.
\end{equation}
Therefore, the bulk field $V$ for the exotic meson has the following
asymptotic form at the boundary $z=\epsilon$:
\begin{equation}
V(z) \;=\; c_1 \frac{1}{z^2} + c_2 z^4,
\label{Asy}
\end{equation}
where $c_1$ is the source term.

We now fix the 5D gauge coupling $\tilde g_5$ by matching the
two-point vector correlation function obtained with the action in
~(\ref{Act}) to the leading contribution from the OPE
result shown in Eq.~(\ref{vOPE})~\cite{Balitsky:1986hf}.  Here we
choose the axial-like gauge
condition $V_z(x,z)\;=\;0$.  It can be decomposed into the transverse
and longitudinal parts: $V_\mu\;=\; (V_\mu)_\perp +(V_\mu)_\|$.
Using the Fourier transform of the vector field: $V_\mu^a=\int d^4x
e^{iq\cdot x} V_\mu^a(x,z)$, we can write the equation of motion for
the transverse part of the vector field as
follows:
\begin{equation}
\left[\partial_z\left(\frac1{z}\partial_z V_\mu^a(q,\,z)\right) +
  \left(\frac{q^2}{z}-C_5^2/z^3 \right) V_\mu^a(q,\,z) \right]_\perp
\;=\;  0\, ,
\label{eq:transv}
\end{equation}
where $C_5^2\equiv m_5^2 \tilde g_5^2$.
The corresponding solution of Eq.(\ref{eq:transv}) can be expressed as
a separable form:
\begin{equation}
(V_\mu^a(q,\,z))_\perp \;=\; V(q,\,z) \overline{V}_\mu^a(q),
\end{equation}
where $\overline{V}_\mu^a (q)$ is the Fourier transform of the source
of the 4D vector-current operator $\bar\psi T^a
G_{\mu\alpha}\gamma^\alpha\psi$.  The explicit solution for $V(q,\,z)$
can be derived by solving Eq.(\ref{eq:transv}),
\begin{equation}
V(Q,z)=c_1zI_{n} (Qz) +c_2 zK_{n}(Qz)\, ,
\end{equation}
where $I_n$ and $K_n$ are modified Bessel functions, $n = \sqrt{1 +
C_5^2}$, and $Q^2=-q^2$.  The asymptotic form of the bulk
vector field at the boundary $z=\epsilon$, which is shown in
Eq.(\ref{Asy}), dictates the following boundary condition:
$V(Q,\epsilon)= c /\epsilon^2$. Here, $c$ is a constant to be
fixed. We refer to ref.~\cite{DaRold:2005vr} for a similar procedure
for a non-exotic scalar two-point correlator.  Imposing the UV
boundary condition $V(Q,\epsilon)\sim 1/\epsilon^2$, we obtain $n=3$
($C_5^2=8$), and $c_2=cK_3^{-1}(Q\epsilon)/\epsilon^3$.  Note that
$C_5^2=8$ implies $\tilde g_5^2=1$ since $m_5^2=8$.  Thus, the
asymptotic behavior of the bulk vector field dictated by the AdS/CFT
correspondence uniquely determines the 5D gauge coupling.  Since the
two-point vector correlation function will be obtained at the UV, 
$z=\epsilon$, we will not explicitly consider the IR boundary
condition that will fix $c_1$.  Then, following the standard procedure
given in \cite{Erlich:2005qh,DaRold:2005zs,DaRold:2005vr}, we obtain
\begin{equation}
\Pi_V(Q^2) \;=\; -\frac1{64}c^2 Q^6 \ln Q^2.
\end{equation}
Comparing this with the OPE given in Eq.~(\ref{vOPE}) in the
leading $Q^2$ order, we obtain
\begin{equation}
c^2=\frac{N_c^2-1}{30\pi^3}\, .\label{c2}
\end{equation}

We are now in a position to predict the mass of $\pi_1$(1400).
Note that we have no free parameter that we may play around to get a
desirable value of  $m_{\pi_1}$.  The mode equation for the exotic
meson reads
\begin{equation}
\left (\partial_z^2-\frac{1}{z}\partial_z+m_i^2-\frac{8}{z^2}
\right)f_i(z)=0\, ,\label{mE}
\end{equation}
where $V_\mu(x,z)=\sum_i f_i(z) V_\mu^{(i)}(x)$.
The solution is given by, in terms of the Bessel functions,
\begin{equation}
f_i(z)=a_1 z J_3 (m_i z) +a_2 z Y_3 (m_i z)\, .
\end{equation}
The KK masses for the exotic mesons are fixed by the IR boundary
condition $\partial_z f_i(z)|_{z=z_m}=0$:
\ba
m_i z_m J_2 (m_iz_m) - 2
J_3 (m_i z_m)=0\, .
\ea
Taking $i=1$ for the ground state, we obtain the
mass of $\pi_1$ to be $m_{\pi_1}\simeq 1476$ MeV that is close to the
experimental value $m_{\pi_1}= 1351\pm 30$ MeV~\cite{Amsler:2008zz}.
\begin{table}[ht]
  \centering
  \begin{tabular}{lcc}
\hline\hline
Models &$\phantom{aaaaaallllllllllllllllllllllll}$ & Mass $[\mathrm{MeV}]$ \\ \hline
Present model & & $1476$ \\
Flux tube
models~\cite{Isgur:1985vy,Close:1994hc,Barnes:1995hc,Page:1998gz,
Swanson:2003kg,Burns:2006wz} & &$1800-2100$ \\
Bag models~\cite{Chanowitz:1982qj,Barnes:1982tx} & & $1300-1800$ \\
QCD sum rules~\cite{Balitsky:1982ps,Balitsky:1986hf,Latorre:1985tg,
Govaerts:1984bk,Govaerts:1984hc,Govaerts:1986pp} &
& $1200-2500$ \\
Lattice QCD\cite{Bernard:2003jd,Hedditch:2005zf,McNeile:2006bz} & & $1800-2300$ \\
Experiment (PDG)~\cite{Amsler:2008zz} & & $1351\pm30$ \\ \hline\hline
  \end{tabular}
  \caption{Comparison with the results of other models}
  \label{tab:1}
\end{table}
In table~\ref{tab:1}, we compare the present result with other
models.  The next excited mass for the hybrid exotic meson turns out
to be $2611$ MeV.  As usual, the excited mass seems to be quite large,
compared to the ground state.

Finally, we consider decay constant of $\pi_1$.  The decay constant of
the $\pi_1$~\cite{Balitsky:1986hf} is defined as
\begin{equation}
\langle 0|\bar{\psi}(x) T^a G_{\mu\alpha}(x)\gamma^\alpha \psi(x)
|\pi_1^b(p)\rangle \;=\; \frac1{\sqrt{2}} F_{\pi_1} m_{\pi_1}^3
\varepsilon_\mu \delta^{ab} e^{ip\cdot x} ,
\end{equation}
where $F_{\pi_1}$ denotes the decay constant of the $\pi_1$. The
$\varepsilon_\mu$ is for its polarization vector. Following the procedure sketched in
  ~\cite{Erlich:2005qh}, we arrive at the definition of $F_{\pi_1}$ in AdS/QCD:
\begin{equation}
(F_{\pi_1} m_{\pi_1}^3)^2 =\frac{c^2}{\tilde g_5^2}\left(\frac{f^\prime_1
    (\epsilon)}{\epsilon^3} \right)^2\, ,
\end{equation}
where $f_1 (z)$ is the normalized solution of the mode equation,
Eq. (\ref{mE}), with $i=1$.
Note that the definition of $F_{\pi_1}^2$ is slightly different from that of
$\rho$-meson given in ~\cite{Erlich:2005qh}:
\ba
F_\rho^2=\frac{1}{g_5^2}[\psi^\prime_\rho(\epsilon)/\epsilon]^2\, .
\ea
The discrepancy is basically due to a difference in UV boundary
conditions: $V(q,\epsilon)=c/\epsilon^2$ for $\pi_1$(1400),
$V(q,\epsilon)=1$ for $\rho$-meson. Taking $N_c=3$ in Eq. (\ref{c2}),
we obtain $F_{\pi_1}=10.6$ MeV.  Compared to the the value of the
decay constant from the QCD sum rules~\cite{Balitsky:1986hf}
$F_{\pi_1}\approx 20$ MeV, the
result is qualitatively in almost the same order. 
\section{Summary}
The present work has aimed at investigating the hybrid exotic mesons
with $J^{PC}=1^{-+}$ within the framework of AdS/QCD.  We  have first
introduced the 5D bulk field dual to the 4D quark-gluon vector
current $\bar{\psi}T^a G_{\mu\alpha}\gamma^\alpha \psi$ and
constructed the 5D bulk action for the exotic meson.
 Solving the
classical equation of motion for the transverse part of the hybrid
exotic vector field, we have obtained the explicit form of the vector
field in terms of the modified Bessel functions with index
$n=\sqrt{1+m_5^2\tilde{g}_5^2}$.  Imposing the UV boundary condition for
the vector field to calculate the two-point correlation function, we
have determined $n=3$, which fixes the 5D gauge coupling.
We have then obtained the eigenvalue equation for the hybrid exotic meson mass.

In order to find the mass of the hybrid exotic meson
$\pi_1(1400)$, we have identified it as the first excited state of the
KK modes.  The mass of the $\pi_1(1400)$ turned out to be
$m_{\pi_1}\simeq 1476\,\mathrm{MeV}$ that is close to the
experimental data: $1351\pm 30\,\mathrm{MeV}$, which
is a remarkable result, considering the fact that the formalism
from AdS/QCD is so simple.  The mass of the next excited
state in the hybrid exotic channel turns out to be $2611$ MeV. Similar to
the $\rho$ mesons in the hard wall model for non-exotic
mesons~\cite{Erlich:2005qh,DaRold:2005zs}, the mass of the excited
state seems quite large in the present study.  We also predicted the
decay constant of $\pi_1$ (1400): 
$F_{\pi_1}=10.6$ MeV. Finally, we remark that it will be
interesting to see if one can study the hybrid exotic meson in a
stringy set-up.

\acknowledgments
We thank Kwanghyun Jo and Sang-Jin Sin for very helpful discussions.
The present work is supported by the Korea Research
Foundation Grant funded by the Korean Government (MOEHRD)
(KRF-2006-312-C00507).

\end{document}